\documentclass{aa}
\usepackage{graphicx}
\usepackage{psfig}
\usepackage{rotating}
\begin{document}

\newcommand{\as}[2]{$#1''\,\hspace{-1.7mm}.\hspace{.1mm}#2$}
\newcommand{\am}[2]{$#1'\,\hspace{-1.7mm}.\hspace{.0mm}#2$}
\def\approxlt{\lower.2em\hbox{$\buildrel < \over \sim$}}
\def\approxgt{\lower.2em\hbox{$\buildrel > \over \sim$}}
\newcommand{\dgr}{\mbox{$^\circ$}}   
\newcommand{\grd}[2]{\mbox{#1\fdg #2}}
\newcommand{\gsim}{\stackrel{>}{_{\sim}}}
\newcommand{\HI}{\mbox{H\,{\sc i}}}
\newcommand{\HIbf}{\mbox{H\hspace{0.155 em}{\footnotesize \bf I}}}
\newcommand{\HIit}{\mbox{H\hspace{0.155 em}{\footnotesize \it I}}}
\newcommand{\HIsl}{\mbox{H\hspace{0.155 em}{\footnotesize \sl I}}}
\newcommand{\HII}{\mbox{H\,{\sc ii}}}
\newcommand{\IHI}{\mbox{${I}_{HI}$}}
\newcommand{\Jykms}{\mbox{Jy~km~s$^{-1}$}}
\newcommand{\kms}{\mbox{km\,s$^{-1}$}}
\newcommand{\kmsMpc}{\mbox{ km\,s$^{-1}$\,Mpc$^{-1}$}}
\newcommand{\nan}{Nan\c{c}ay}\def\lir{{\hbox {$L_{IR}$}}}
\def\lco{{\hbox {$L_{CO}$}}}\def \ls{\hbox{$L_{\odot}$}}
\newcommand{\LB}{\mbox{$L_{B}$}}
\newcommand{\LBnul}{\mbox{$L_{B}^0$}}
\newcommand{\LBsun}{\mbox{$L_{\odot,B}$}}
\newcommand{\lsim}{\stackrel{<}{_{\sim}}}
\newcommand{\LsunK}{\mbox{$L_{\odot, K}$}}
\newcommand{\LsunB}{\mbox{$L_{\odot, B}$}}
\newcommand{\LsunMsun}{\mbox{$L_{\odot}$/${M}_{\odot}$}}
\newcommand{\LK}{\mbox{$L_K$}}
\newcommand{\LKLB}{\mbox{$L_K$/$L_B$}}
\newcommand{\LKLBnul}{\mbox{$L_K$/$L_{B}^0$}}
\newcommand{\LKLsun}{\mbox{$L_{K}$/$L_{\odot,Bol}$}}
\newcommand{\masq}{\mbox{mag~arcsec$^{-2}$}}
\newcommand{\MHI}{\mbox{${M}_{HI}$}}
\newcommand{\MHILB}{\mbox{$M_{HI}/L_B$}}
\newcommand{\MHILBfr}{\mbox{$\frac{{M}_{HI}}{L_{B}}$}}
\newcommand{\MHILK}{\mbox{$M_{HI}/L_K$}}
\newcommand{\MHILKfr}{\mbox{$\frac{{M}_{HI}}{L_{K}}$}}
\def \ms{\hbox{$M_{\odot}$}}
\newcommand{\Msun}{\mbox{${M}_\odot$}}
\newcommand{\MsunLsun}{\mbox{${M}_{\odot}$/$L_{\odot,Bol}$}}
\newcommand{\MsunLBsun}{\mbox{${M}_{\odot}$/$L_{\odot,B}$}}
\newcommand{\MsunLKsun}{\mbox{${M}_{\odot}$/$L_{\odot,K}$}}
\newcommand{\MT}{\mbox{${M}_{ T}$}}
\newcommand{\MTLBnul}{\mbox{${M}_{T}$/$L_{B}^0$}}
\newcommand{\MTLBsun}{\mbox{${M}_{T}$/$L_{\odot,B}$}}
\newcommand{\tis}[2]{$#1^{s}\,\hspace{-1.7mm}.\hspace{.1mm}#2$}
\newcommand{\Vcor}{\mbox{${V}_{0}$}}
\newcommand{\vhel}{\mbox{$V_{hel}$}}
\newcommand{\VHI}{\mbox{${V}_{HI}$}}
\newcommand{\vrot}{\mbox{$v_{rot}$}}
\def\la{\mathrel{\hbox{\rlap{\hbox{\lower4pt\hbox{$\sim$}}}\hbox{$<$}}}}
\def\ga{\mathrel{\hbox{\rlap{\hbox{\lower4pt\hbox{$\sim$}}}\hbox{$>$}}}} 

\title{A search for Low Surface Brightness galaxies in the near-infrared}
  \subtitle{I. Selection of the sample}  

\author{D. Monnier Ragaigne\inst{1},           
        W. van Driel\inst{1},          
        S.E. Schneider\inst{2},	  
        T.H. Jarrett\inst{3}       
       \and          
        C. Balkowski\inst{1}          
         }   

\offprints{W. van Driel}  
\institute{Observatoire de Paris, GEPI, CNRS UMR 8111 and Universit\'e Paris 7, 
           5 place Jules Janssen, F-92195 Meudon Cedex, France \\
            \email{delphine.ragaigne@obspm.fr; wim.vandriel@obspm.fr; 
                   chantal.balkowski@obspm.fr}
           \and             
               University of Massachusetts, Astronomy Program, 536 LGRC, Amherst, 
               MA 01003, U.S.A. \\
             \email{schneide@messier.astro.umass.edu}
           \and             
               IPAC, Caltech, MS 100-22, 770 South Wilson Ave., Pasadena,              
               CA 91125, U.S.A. \\             
             \email{jarrett@ipac.caltech.edu}             
             }     
\date{\it Received 14/2/2002 ; accepted 22/4/2002}

\abstract{ {\rm 
A sample of about 3,800 Low Surface Brightness (LSB) galaxies was selected using
the all-sky near-infrared ($J$, $H$ and $K_s$-band) 2MASS survey. The selected 
objects have a mean central surface brightness within a 5$''$ radius around 
their centre fainter than 18 \masq\ in the $K_s$ band, making them the lowest
surface brightness galaxies detected by 2MASS.  
A description is given of the relevant properties of the 2MASS 
survey and the LSB galaxy selection procedure, as well as of basic 
photometric properties of the selected objects. The latter properties 
are compared to those of other samples of galaxies, of both LSBs and 
`classical' high surface brightness (HSB) objects, which were selected 
in the optical. The 2MASS LSBs have a $B_{T_{c}}-K_T$ colour which is on 
average 0.9 mag bluer than that of HSBs from the NGC. The 2MASS sample 
does not appear to contain a significant population of red objects.
}
} 

\authorrunning{D. Monnier Ragaigne et al.} 
\titlerunning{A search for LSB galaxies in the near-infrared I.} 
\maketitle

\section{Introduction}   
Observational bias in the selection of galaxies dates back to the visual searches
by Messier and Herschel, as galaxies are diffuse objects selected in the presence 
of a contaminating signal: the brightness of the night sky. 
The night sky acts as a filter, which, when 
convolved with the true population of galaxies gives the population of galaxies 
we observe and which hampers the detection of low surface brightness objects. 
Based on this concern, the result of the study of Freeman (1970), which
indicated a remarkably narrow range of central disc surface brightness values of
spiral galaxies in the $B$-band (21.7$\pm$0.3 \masq), was questioned by Disney (1976).

In the past few decades, observations of the local universe have 
shown the existence of galaxies well below the surface brightness of the average 
previously catalogued galaxies, and often even well below the surface brightness 
of the night sky, which are referred to as Low Surface Brightness (LSB) galaxies.

At present, the LSBs constitute the least well known fraction of galaxies: their 
number density and physical properties (like luminosity, colours, dynamics) are still 
quite uncertain. This is, per definition, mainly due to the fundamental difficulty 
in identifying them in imaging surveys and in measuring their properties. 
It is far from obvious that all existing classes and types of LSBs have been probed, 
given the limitations in sensitivity of the actual detectors and the biases in
the selection criteria of the published samples.

In order to further investigate the often baffling properties of the LSB class of 
galaxies we selected a large sample of them from the 2MASS database, accessing 
the near-infrared, a wavelength domain that has been much less fully explored 
than the optical in the study of LSBs.

To date, the following studies were published on the near-infrared properties
of LSB galaxies: Knezek \& Wroten (1994) observed a small sample of mainly
massive galaxies in $JHK$, Bergvall et al. (1999) observed 14 blue LSBs in
$JHK$ and compared their results to optical photometry, Bell et al. (2000) 
observed 26 LSBs in $JHK$ and a subset of these in $BV\!R$, de Jong (1996) included
a few LSBs among the spiral galaxies he observed in $H$ and $K$, and
Galaz et al. (2002) observed 88 LSBs in $J$ and $K_s$.

Although there is no unambiguous definition of LSB galaxy, those in common use
are based on (1) the mean blue surface brightness within the 25 \masq\ isophote, 
(2) the mean surface brightness within the half-light radius, or (3) the 
extrapolated central surface brightness of the disc component alone, after 
carrying out a disc-bulge decomposition.

For 2MASS galaxies, we used the mean $K_s$-band magnitude within a fixed 
aperture to identify a sample of galaxies with relatively low infrared 
surface brightness. We selected galaxies in which the central surface brightness 
within a 5$''$ radius circular aperture was fainter than 18 \masq\ in the 
$K_s$-band. This criterion selects a relatively small fraction ($<10\%$) of large
angular diameter sources detected by 2MASS, increasing to about 40\% for
the small angular diameter sources detected in the survey. 

If 2MASS detections have $B-K=4.0$ colors typical of early-type galaxies (Jarrett et al.
2003; also see Section 4.4), as might be anticipated for a sample of infrared-selected
sources, this criterion would correspond roughly to the disc-component 
definition that LSB galaxies are objects with a blue central surface brightness 
$\mu_{B_0}>22.0$ \masq (i.e., more than 1$\sigma$ fainter than the 21.7$\pm$0.3 \masq\
measured by Freeman 1970). 
Our sample galaxies can have an even lower disc surface brightness if the bulge 
component is significant, but since we average over a fixed angular aperture, on 
the other hand we may also include some higher central surface brightness sources 
that are more distant so that the aperture includes 
more of the disc. In this paper, we examine the properties of our sample based 
on 2MASS and LEDA data and compare it to optically-selected galaxy samples.

LSBs have remarkable properties which distinguish them from 'classical' 
High Surface Brightness (HSB) spirals, notably:
\begin{itemize}
\item LSBs seem to constitute at least 50\% of the total galaxy population in 
number in the local Universe, which may have  strong implications for the 
faint end slope of the galaxy
luminosity function, on the baryonic matter density and especially on galaxy 
formation scenarios (O'Neil \& Bothun 2000).
\item LSBs discs are among the less evolved objects in the local universe since 
they have a very low star formation rate (van der Hulst et al. 1993; 
van Zee et al. 1997; van den Hoek et al. 2000).
\item LSBs are embedded in dark matter halos which are of lower density and more 
extended than the haloes around HSB galaxies, and they appear to be strongly 
dominated by Dark Matter at all radii (e.g., de Blok et al. 1996, 2001;
McGaugh et al. 2001)
\end{itemize}

The star formation history of LSBs has been the subject of recent debate.
The LSBs best studied in the optical and in the near-infrared are blue (e.g., 
Bergvall et al. 1999), indicating a young mean stellar age and/or metallicity. 
Morphologically, most studied LSBs have discs, but little spiral structure.
The current massive star formation rates in LSBs are an order of magnitude 
lower than those of HSBs (van Zee et al. 1997); \HI\ observations show that 
LSBs have high gas mass fractions, sometimes exceeding unity 
(Spitzak \& Schneider 1998; McGaugh \& de Blok 1997). All these observations 
are consistent with a scenario in which LSBs are relatively unevolved, 
low mass surface density, low metallicity systems with roughly constant 
star formation rate. However, this scenario has some difficulty accommodating 
giant LSBs like Malin 1 (Bothun et al. 1987), see, e.g., Boissier et al. (2003).

This study of infrared LSBs was also intended to investigate the possibility
of there being a substantial population of red LSBs like those reported in 
the optical $UBV\!RI$-band study of O'Neil et al.~(1997). Follow-up \HI\ line studies
initially indicated that some of these red LSBs with rotational speeds
exceeding 200 \kms\ did not seem to follow the `standard' Tully-Fisher relation 
established for HSB galaxies, in the sense that they appeared to be severely 
underluminous for their total mass (O'Neil et al.~2000). More recent 
\HI\ imaging observations by Chung et al.~(2002) indicate that the rotational 
properties and total \HI\ mass of these red LSBs had been strongly overestimated 
due to confusion within the telescope beam with neighbouring galaxies. 
An infrared-selected sample should allow us to identify whether there is a significant 
population of very red LSBs.

In the present paper a description is given of the 2MASS LSB galaxy sample 
selection and its results, while 21-cm \HI\ line observations from Arecibo 
and \nan\ will be presented in papers II and III, respectively 
(Monnier Ragaigne et al. 2003b,c), optical $BV\!RI$-band CCD surface photometry 
of a sub-sample of 35 2MASS LSB galaxies will be presented in paper IV 
(Monnier Ragaigne et al. 2003d) and an analysis of the
full data set will be presented in paper V.
Models of the evolution of the 2MASS galaxies presented in paper IV and of 
other samples of LSB galaxies are presented in Boissier et al. (2003).

The 2MASS survey is described in Section 2, and the 2MASS LSB galaxies sample 
selection in Section 3. Basic photometric properties of the sample are presented 
in Section 4 and compared with those of others samples of LSB and HSB galaxies.
Subsets of the sample observed by us at other wavelengths are described in 
Section 5.

\section{The 2MASS all-sky near-infrared survey}  
The Two Micron All Sky Survey, 2MASS,
has imaged the entire celestial sphere in the near-infrared $J$ (1.25$\mu$m), 
$H$ (1.65$\mu$m) and $K_s$ (2.16$\mu$m) bands using two identical dedicated 
1.3-meter telescopes, designed specifically for the survey. One goal of the 
survey was to probe the nearby Universe in detail out to redshifts of $z\sim 0.1$. 
Accordingly, it is highly uniform in both photometric and astrometric 
measurements. The 2MASS Extended Source Catalog (XSC) consists of more than 1.6 
million galaxies brighter than 14th mag at $K_s$ with angular diameters greater 
than $\sim$10$''$. The photometry includes accurate Point Spread Function 
(PSF)-derived measures and a variety of circular and elliptical aperture measures, 
fully characterizing both point-like and extended objects. The position centroids 
have an astrometric accuracy better than $\sim$\as{0}{5}. In addition to tabular 
information, 2MASS archives full-resolution images for each extended object, 
enabling detailed comparison with other imaging surveys. 

Though 2MASS perforce needed to sacrifice depth for all-sky coverage, and is 
therefore less deep than some of the dedicated optical imaging surveys made 
of LSB galaxies over limited areas of the sky, it should still be sensitive 
to the bright end of the LSB spectrum, particularly if there is a class of
very red LSB objects. It has a 95\% completeness level in 
$J$, $H$ and $K_s$ of 15.1, 14.3 and 13.5 mag, respectively, for `normal' 
galaxies (Jarrett et al. 2000); for LSB and blue objects the completeness 
limits are not yet known. In practice, 2MASS detects galaxies with central surface 
brightness values ranging from 14-20 \masq\ of which the LSB fraction with 
$18<\mu_{K5}<20$ \masq\ varies from $\sim 10$\% of the total 2MASS sample 
for large galaxies ($r_{K_{20}} \geq20''$) to $\sim 40$\%  for the smaller 
galaxies ($20'' \geq r_{K_{20}} \geq 10''$). These near-infrared data will be 
less susceptible than optical surveys to the effects of extinction due to 
dust, both Galactic and internal to the galaxies. Also, the luminous mass 
distribution of galaxies is dominated by the older stellar 
populations, which emit most of their light in the near-infrared.

Initial results for galaxies detected by 2MASS, including LSB objects, are 
described in several publications (Schneider et al. 1997; Jarrett et al. 2000a,b; 
Hurt et al. 2000).

\subsection{The Extended Source Processor}  
The last major subsystem to run in the 2MASS quasi-linear data reduction 
`pipeline' is the extended source processor, GALWORKS. The primary role 
of the processor is to characterize each detected source and decide which 
sources are extended or resolved with respect to the point spread function 
(PSF). Sources that are deemed extended are measured further and the 
information is written into a separate table. In addition to tabulated 
source information, a small ``postage stamp'' image is extracted for each 
extended source from the corresponding $J$, $H$ and $K_s$ Atlas images. 
The source lists and image data are stored in the 2MASS extended source 
database. By the time GALWORKS is run in the 2MASS pipeline, point sources 
have been fully measured with refined positions and photometry, band-merged, 
coordinate positions 
calibrated, Atlas images constructed, and the time-dependent PSF 
characterized for every Atlas image. The high-level steps that encompass 
GALWORKS include: 
(1) bright star (and their associated features) masking, (2) Atlas image 
background subtraction, (3) measurement of the stellar number density and 
confusion noise, (4) source parameterisation and attribute measurements, 
(5) star-galaxy discrimination, (6) refined photometric measurements, and 
finally (7) source and image extraction (see Jarrett et al. 2000a for more details). 

The sample selected for this study was extracted before the 2MASS data 
acquisition was completed, using an earlier version of the software extraction
algorithms. However, comparisons between the final data and the earlier
results show they are consistent. We also used the full 2MASS database, not
just the high signal-to-noise sources that form the final catalog, so that
our sample includes many fainter sources than are part of the publicly released
catalogs.

\subsection{Low Central Surface Brightness (LCSB) Source  Processor}  
There are some galaxies whose central surface brightness is too low to be 
detected by the standard 2MASS procedure, but whose total integrated flux 
is significant. These may include low surface brightness galaxies and dwarf 
galaxies. We will refer to these sources with the generic moniker: 
Low Central Surface Brightness (LCSB) galaxies. LCSB galaxies present 
a different challenge to GALWORKS than the typical `normal' galaxy 
2MASS encounters. They are generally very faint (as measured in a standard 
aperture for `normal' galaxies) and they do not have well defined cores; 
see Jarrett (1998) for examples. The LCSB processor was an experimental 
algorithm executed last in the 
chain of operations that comprise GALWORKS. Sources detected by this
algorithm did not meet the stringent reliability criteria for inclusion
in the public 2MASS catalogue, but it did occasionally identify interesting
LSB galaxies, as confirmed with deep optical images. There are also some examples 
of galaxies observed to be low surface brightness in the near-infrared that 
have normal surface brightness in the optical, see, e.g., Galaz et al. (2002). 

The input to the LCSB processor is a fully cleaned Atlas image in 
each band, where stars and previously found extended sources have been 
entirely masked. The image is then blocked up (using three independent 
kernel sizes: $2\times2$, $4\times4$ and $8\times8$ pixels) and `boxcar' 
smoothed to increase the signal-to-noise ratio for large (but faint) 
objects normally hidden in the pixel noise. The detection step consists of 
3-$\sigma$ threshold isolation of local peaks in cleaned, smoothed images. 
Source detections are then parameterised, with the primary measurements 
being: signal-to-noise ratio of the peak pixel, radial extent, 
integrated signal-to-noise ratio, surface brightness, integrated 
flux, and signal-to-noise ratio measurements using a $J$+$H$+$K_s$ combined 
`super-coadd' image. In principle, the `super-coadd' provides the best 
medium from which to find faint LSB galaxies, given the effective increase 
in the signal-to-noise ratio.

The LCSB processor could be falsely triggered by faint stars and the wings
of the PSF around bright stars. Meteor streaks (and other transient phenomena) 
that were not fully cleaned from the Atlas images could also generate
numerous false LCSB sources. 
It is important to note that LCSB-detected sources were nearly always fainter than the 
catalogue limit of the 2MASS survey, so their exclusion from the released
catalogue does not significantly compromise its completeness. 
Further information and some early science results with 
2MASS LCSB galaxies can be found in Jarrett et al. (2000a).

\section{2MASS LSB galaxies selection}  
After a period of testing with various combinations of photometric parameters, 
we defined a {\it modus operandi} for the sample selection based on as few 2MASS 
parameters as possible, which allowed the definition of a sample with low 
infrared surface brightness values. The 2MASS central surface brightness limit 
of $\sim$20 \masq\ in $K_s$ band is equivalent to a B surface brightness of 
$\sim$24 \masq\ for a typical spiral galaxy colour. This corresponds to fairly 
`bright LSB' galaxies, easily detected in deep photographic and CCD optical 
limited-coverage surveys. However, as we shall show later in this paper, 
our selection criteria yield a LSB sample with interesting complementary 
and contrasting properties compared to optically-selected samples. 

\subsection{Selection using the $K_s$-band central surface brightness}  
In order to select infrared LSB galaxies, our key photometric 2MASS parameter 
is $\mu_{K5}$, the mean $K_s$-band central surface brightness within a 
circle of radius 5 arcsec. This aperture was selected because for the great 
majority of faint galaxies in the 2MASS catalogue, this small fixed circular 
aperture gives good signal-to-noise statistics while avoiding problems due 
to confusion and missing flux in the faint outer parts of galaxies. 
The circular 5$''$ radius aperture also appears more reliable than the peak 
central surface brightness measured in a single pixel of the original 
1.8 sec integration.

We selected sources with $\mu_{K5}>18.0\, \hbox{\masq}$. The usefulness 
of this criterion for the selection of LSB galaxies has been shown in 
Jarrett et al.~(1998) for selecting LSB galaxies in the Coma cluster and SA57 region.

\subsection{Selection using source size}  
The size of the sources is also important for the sample selection:

\begin{itemize}
\item A fixed-diameter aperture gives the surface brightness over different 
physical areas within the galaxy, depending on its distance. For angularly 
small galaxies, the fixed aperture gives a surface brightness closer to the 
mean over the whole galaxy.
\item The reliability of the catalogue is related to the size of the objects. 
For very small sources, confusion with stars increases significantly
\item We planned also to obtain observation in the 21 cm \HI\ radio line, 
and flux levels tend to be correlated with angular size.
\end{itemize}

For these reasons, we selected two samples of galaxies based on $r_{K_{20}}$, the 
$K_s$-band semi-major axis at an isophotal level of 20 \masq. Our primary sample 
is of sources with $r_{K_{20}}>20''$ (hereafter referred to as the `Large' 2MASS  
LSBs sample), and a secondary sample with  $20''>r_{K_{20}}>10''$ (hereafter referred 
to as the `Small' 2MASS LSBs sample). 

\subsection{Selection using colour range}  
Two effects make galaxies appear `redder' in the 1 to 2 $\mu$m window (see Figure 
14, Section 5, in the Explanatory Supplement to the 2MASS Second Incremental Data):
\begin{itemize}
\item their light is dominated by older stellar populations
\item their redshift tends to transfer additional stellar light into the 2$\mu$m 
windows, boosting the $K_s$ band flux relative to the $J$ band flux 
(K-correction)
\end{itemize}
In general, the $(J-K_s)$ colour falls within a fairly narrow range for nearby 
galaxies, $0.8<(J-K_s)<1.3$, and GALWORKS uses this fact to help discriminate 
galaxies from other Galactic extended sources. Because we are exploring a 
potentially unusual class of galaxies, we examined a wider colour range of 
sources from the ''Working Survey Database'' of 2MASS (including uncertain 
sources that do not meet the completeness and reliability criteria for the 
2MASS catalogue). To exclude most artifacts and confusing sources, we select 
sources with $0.5<(J-K_s)<2.0$. This range also allows for larger photometric 
uncertainties that are more likely to be found for LSB galaxies. 

\subsection{Sky coverage}   
The presently available 2MASS database covers all the sky. 
At the time when the sample was selected (end of 1999) the declination
range of $\sim$0$^{\circ}$ to +12$^{\circ}$ had not yet been covered, however, 
as can been seen in Figure 1, which shows the sky coverage of the objects from 
our sample observed in the \HI\ line.

To select our sample of LSB galaxies, we exclude the region inside the Zone of 
Avoidance ($|$b$|$$<$10$^{\circ}$) where the confusion rate between multiple 
stars and galaxies is high, as well as the declinations not suited for observations 
with the Nan\c{c}ay and Arecibo radio telescopes.

Thus the area of the sky covered by our survey ranges from -39$^{\circ}$ to 
$\sim$0$^{\circ}$ and from $\sim$12$^{\circ}$ to +60$^{\circ}$ in declination, 
and the galactic latitude of our sources is $|$b$|$$>$10$^{\circ}$.

\subsection{Final sample}  
Using the first and second algorithms described in Sections 2.1 and 2.2, we 
selected, respectively, 34,000 and 318 LCSB sources with a mean central 
surface brightness $\mu_{K5}$ fainter than 18 \masq. All sources have 
a radius $r_{K_{20}}$ larger than 10 arcsec.

In order to decide which of these faint sources really are galaxies, 
additional data on them were used that were listed in online databases 
such as NED (NASA Extragalactic Database) [http://nedwww.ipac.caltech.edu], 
LEDA (Lyon-Meudon Extragalactic Database)  -- recently incorporated in HyperLeda 
[http://www-obs.univ-lyon1/hypercat/] -- and Aladin of the Centre de Donn\'ees 
astronomiques de Strasbourg (CDS) [http://aladin.u-strasbg.fr] , 
and by inspecting their Digital Sky Survey (DSS) optical images. 
As the acquisition and interpretation of these supplementary data for 34,000 
sources was deemed to be too laborious, and as we intended, after a first 
inspection of the sample, to concentrate our multi-wavelength observations 
(see Section 5) on the larger-sized objects, we limited this phase of the 
selection procedure to all 2,000 objects in the subset of `large' sources 
($r_{K_{20}} > 20''$) and to 3,000 randomly selected objects in the subset of 
`small' sources ($20''> r_{K_{20}} > 10''$). Using this procedure, we selected 
a total of 3,736 and 59 candidate 2MASS LSB galaxies from the sources 
extracted by the GALWORKS standard and LCSB algorithms, respectively. 
The total sample of 3,796 objects constitutes our working database of 
2MASS LSB galaxies.

As the selected LCSB sources are not part of the standard 2MASS data released for
public use, unlike the other sources, we show their near-infrared images in Figure 2.

\section{Basic properties of the sample }  
Some basic parameters of our working database of 3,795 2MASS LSB galaxies 
are listed in Tables 1, 2 and 3. We have divided the subsets listed in the 
Tables according to two criteria: processor and size. Tables 1 and 2 list 
the objects extracted using the standard algorithm for, respectively, the 
Large and Small sources, while Table 3 lists the Faint objects extracted using 
the LCSB algorithm.

It should be noted that the selection of the infrared sources presented in this work,
which are all derived from 2MASS $J$, $H$ and $K_s$-band imaging, was made in
late 1999, when work on the 2MASS database was still in full progress.  
The five-year evolution of both detection and source
characterization was driven by improvements in the pipeline reductions
and the image calibration.  These are realized through preliminary database
holdings and public ``sampler'' data releases, incremental data releases and finally
the All Sky data release that achieved its milestone in March of 2003. 
Accordingly, the current sample is a mixture of these data releases,
which are described in the 2MASS Explanatory Supplement, 
see URL http://www.ipac.caltech.edu/2mass/releases/\-allsky/doc/explsup.html.

The Large and Small sources put in parentheses in Tables 1 and 2, respectively, 
are those that are only found in the All Sky working databases, and not in the subsequent, 
more reliable, public data releases. 
Only 23 of the 836 Large sources we selected from the working
databases do not occur in later data releases, as do 27 from the 2900 Small sources.
Of these 50 non-confirmed sources, 41 have an entry in the PGC and 24 in other optical
galaxy catalogues. 

The situation is different for the 59 Faint sources, as these were selected 
for our survey using a dedicated LSB source detection routine (see Section 2.2), and none 
of them appeared in the working databases. 
Seventeen of them do appear in later public releases, however: see Table 3, where, 
for the sake of consistency, we have also put
the names of the Faint sources that are not found in later public data releases in brackets.

The basic parameters listed in Tables 1, 2 and 3 are as follows:

\begin{enumerate}
\item No: the `work name', used throughout the present series of papers, 
which consists of  the type of the source (L: large; S: Small; F: LCSB
(Faint source) processor) followed by its number and an indication of 
its inclusion in other catalogues (P: only in PGC (see Paturel et al. 1989, 
as updated continuously in LEDA), O: in other catalogue(s), N: not previously 
catalogued). Sources in parentheses did not appear in later public data releases
(see above);
\item 2MASS: gives the J2000.0 right ascension and declination coordinates 
of the 2MASS source;
\item PGC:  the PGC catalogue entry, if available;
\item Other: refers to entries in other galaxy catalogues;
\item $\mu_{K5}$: the mean central surface brightness (in \masq) measured
within a radius of 5 arcsec around the source's centre in the $K_s$ band;
\item $\mu_{B_{25}}$: the mean $B$-band surface brightness (in \masq)
within the $D_{25}$ blue isophotal diameter, at a level of 25 \masq, as 
listed in LEDA;
\item Obs: information on available observations: A for our Arecibo \HI\ 
detections and `a' for the non-detections (see paper II), N for our \nan\ \HI\ 
detections and `n' for the non-detections (see paper III), O for other published 
\HI\ detections (from LEDA, see paper V), and P for our $BV\!RI$ optical surface 
photometry data (see Paper IV).
\end{enumerate}

\subsection{2MASS and comparison samples}  
To study the basic properties of our galaxy sample, we compared various subsets 
to existing samples of optically selected High and Low Surface Brightness galaxies. 
We compared near-infrared and optical data from LEDA for the objects in the 
following samples which were detected in 2MASS:
\begin{enumerate}
\item 2MASS L: our `Large' LSB galaxies from the 2MASS survey catalogue with a
 20 \masq\ $K_s$-band isophotal radius $r_{K_{20}} \geq 20''$ [836 objects]; 
\item 2MASS S: same for the `Small' objects, with $20'' \geq r_{K_{20}} \geq 10''$ [2900 objects]; 
\item 2MASS F: 2MASS sources with lower central surface brightness, extracted 
using the LCSB algorithm (see Sect. 2.2) [59 objects]; 
\item NGC L: `classical' HSB galaxies from the NGC,  with $r_{K_{20}} \geq 20''$ 
[1973 objects]; 
\item NGC S: same, with $r_{K_{20}} \leq 20''$ [252 objects]; 
\item LSB opt.: optically selected LSB galaxies from Impey et al. (1996) [83 objects].
\end{enumerate}

For the last sample of 693 optically selected LSB galaxies (derived from the 
inspection of UK Schmidt photographic plates) 83 objects were found in the 2MASS 
working survey database (i.e., 12\%). These LSBs found in the 2MASS database 
have a central surface brightness in $B$ (measured as described by 
Sprayberry et al.~1996) between 18.2 and 25.0 \masq\ with a mean value 
around 21.6 \masq\ [i.e., the Freeman value for HSB spiral discs] compared to 
the other objects in the Impey et al. sample, which have a median value 
around 22.8 \masq.

\subsection{Infrared surface brightness in the $K_s$ band}  
In Figure 3 we compare several different measures of the $K_s$-band surface 
brightness for the subsets of the 2MASS LSB sample and the optically selected 
samples:\\
{\sf $\bullet$ Central $K_s$-band surface brightness:}\, $\mu_{K5}$ is the 
primary selection criterion for inclusion in our 2MASS LSBs sample, the mean surface 
brightness measured within a circle of 5$''$ radius at the centre of each galaxy. 
Both the Large and Small 2MASS samples exhibit the highest counts at the 
18 \masq\ cut-off, while the sources detected with the LCSB algorithm peak 
about 1 \masq\ lower. The 2MASS-selected sources span the lower surface 
brightness range of the optically selected LSBs detected by 2MASS. 
By contrast, the NGC sample objects are, on average, about 1.5 to 2 \masq\ 
brighter than the Small and Large 2MASS samples, respectively. The difference 
between the two NGC samples appears to be due to the inclusion of more of the 
disc in the small galaxies within the 5$''$ radius aperture. \\
{\sf $\bullet$ Peak $K_s$-band surface brightness:}\, $\mu_0$ is the highest 
surface brightness measured in any of the 2$''$ camera pixels near the centre 
of the galaxy during the original observations. 2MASS made six observations 
in a dithered pattern so that one observation was likely to be well-centred 
on the galaxy nucleus and should normally have the highest surface brightness. 
The results for the various samples are approximately the same as for the central 
surface brightness, although the Large and Small 2MASS samples become better 
distinguished. The small galaxies have slightly higher peak surface brightness 
values on average because the $\sim1''$ radius aperture is more dominated by 
the bulge than the $5''$ aperture, whereas the measurements 
of the larger galaxies were already more bulge dominated. 
\\
{\sf $\bullet$ Mean isophotal surface brightness} and {\sf mean face-on 
isophotal surface brightness:}\, $\mu_{K_{21}}$ and $\mu_{K_{21}} (i=0)$ are, 
respectively, the average surface brightness measured within the ellipse 
fit to the $K_s$-band 21 \masq\ isophotal contour and this value corrected 
to face-on $(i=0^{\circ})$. The latter value simply uses the major axis diameter 
as a measure of the face-on radius. These definitions corresponds to older, 
optical, definitions of surface brightness that used the $B=25$ \masq\ 
isophotal size, $D_{25}$. Note again that our 2MASS samples exhibit some of the 
lowest infrared surface brightness values of any of the samples. 
As this Figure shows, especially the Large sample is expected to contain a
relatively many highly inclined, intrinsically quite LSB 
sources, due to selection effects (see Section 4.3).
The optical LSB sample has relatively high surface brightness levels under 
this measurement scheme, because these galaxies may have a bright bulge even 
though their disc component has a low surface brightness,
and a significant fraction may even have relatively bright discs,
as the Impey et al. sample contains a significant number of non-LSB galaxies.

\subsection{Other infrared properties}  
We compare a few other basic properties of the infrared emission in Figure 4:

\noindent
{\sf $\bullet$ $K_s$-band isophotal radius:}\, $r_{K_{20}}$ is the major-axis 
radius at the $K_s=20$ \masq\ elliptical isophote. Our 2MASS samples tend 
to be dominated by sources near the minimum radius, while the NGC samples are 
weighted toward angularly larger objects. The optical LSB sample is also dominated 
by smaller sources. \\
{\sf $\bullet$ $K_s$-band isophotal magnitude:}\, $K_{20}$ is the total magnitude 
within the $K_s$=20 \masq\ elliptical isophote. Here again the 2MASS and optical 
LSB samples are weighted toward sources nearer the completeness limits of 2MASS, 
while the NGC samples include relatively bright objects. \\
{\sf $\bullet$ Infrared axis ratio:}\, $b/a$ is determined from an ellipse fit 
to the co-addition of the $J$-, $H$-, and $K_s$-band 
images. The fit is made at the 3-$\sigma$ isophotal level relative to the 
background noise in each image. (The Faint 2MASS LCSBs sample was not measured 
in this way because of the low S/N levels of the emission.) A sample of randomly 
oriented disc galaxies would exhibit a relatively flat distribution of $b/a$, 
while inclusion of elliptical galaxies and systems dominated by the bulge component 
tends to give a rising distribution toward larger values of $b/a$ 
(e.g., Mihalas \& Binney 1981). The NGC and optical LSB samples exhibit 
some tendency toward the distribution expected for bulge domination in 
the infrared, but the 2MASS samples show dominance by high-inclination systems,
which is more pronounced for the Large sample than for the Small sources.
This is due to  the combined effects of the limits in observed central surface brightness 
and in apparent diameter, since, measured at a given isophotal level,  highly inclined galaxies 
appear larger than more face-on objects with a similar intrinsic surface brightness 
profile. In particular the Large sample will contain an
over-representation of highly inclined, intrinsically quite low surface brightness sources. \\
{\sf $\bullet$ Infrared colour:}\, $(J-K)_{20}$ is the difference in $J$- and 
$K_s$-band magnitudes measured within the $K_s=20$ \masq\ elliptical isophote. 
The samples peak near $(J-K)=1$, which is found quite generally for all galaxy 
types. The LCSB sample is bluer (0.9 mag), while the Small 2MASS sample is somewhat 
redder (1.2 mag) than the average. Although these colour differences may 
partly reflect differences in the K-correction due to galaxy redshift 
(bluer for nearer, redder for more distant), these are estimated to be of the
order of 0.1 mag at most, however, and cannot explain the much larger  observed 
differences.

\subsection{Optical-infrared colours}  
In Figure 5 we compare optical and optical-infrared colours of the 2MASS and optical 
samples. We show several combinations based on total corrected
extrapolated magnitudes from 2MASS measured in the $K_s$ band and from LEDA measured 
in the $U$, $B$, $V$ and $I$ bands. 

The total corrected $K_s$-band infrared magnitudes, $K_T$, were derived by modelling 
the increase in flux with successively larger apertures and extrapolating to a total 
magnitude. No corrections were applied for inclination or extinction, as these are 
in practice negligible at this wavelength.
The estimated uncertainty in the extrapolated $K_T$ magnitudes is of the
order of 0.15 mag for $K_T$$\sim$13 mag, a typical value for the sample.

The total corrected optical magnitudes from LEDA are mostly derived from isophotal 
magnitudes, with corrections for galactic extinction, inclination and extinction 
effects (see Paturel et al. 1997). As the extinction correction depends on the 
morphological type, only galaxies with a $B_{T_{c}}$ value and a morphological 
type listed in LEDA were considered. This limits the percentage of useable 2MASS 
LSB galaxies in the Large, Small and Faint categories to 63\%, 15\% and 5\%, 
respectively, and the optically selected LSBs to 60\%.

In comparisons between the $U$, $B$, $V$, and $I$ bands, 
both the Large and Small 2MASS samples tend to be 
bluer than the optical samples: the optical LSB sample (``LSB opt'' in plot)
is next bluest, and the NGC samples exhibit the reddest colours. (The LCSB Faint source sample had too 
few sources with LEDA optical magnitudes to show meaningful colour distributions 
in these plots.) It is well established that the average HSB galaxy tend to be redder than 
the average, non-dwarf LSB (e.g., O'Neil et al. 1997; Gerritsen \&  de Blok 1999; Burkholder et al. 2001).
The presence in the Impey et al. ``optical LSB'' sample of a significant 
fraction of galaxies with relatively high central surface brightness, which can in fact be 
considered as HSBs, may well explain the intermediairy colours found for this sample.

Regarding the $(B_{T_{c}}-K_{T})$ colour, the Large and Small NGC galaxies have 
an average of 3.6 and standard deviations of 1.3 and 0.3, respectively, while 
the optical LSBs are slightly bluer, 3.2$\pm$0.8. Of the 2MASS samples, that 
of the Faint objects is the reddest on average, 3.2$\pm$0.6, while the Small 
sources are also fairly red, 3.0$\pm$1.1, but the mean of the Large 2MASS sample 
is 2.7$\pm$0.9. Thus the infrared LSBs are not necessarily optical LSBs, according 
to our working definition of $\mu_{B_0,d}$$>$22 \masq, as the adopted $K_s$-band 
selection limit of 18 \masq\ corresponds to a mean blue value of $\sim$20.8 \masq\ 
for their mean colour, though this includes the (unknown) contribution from the 
bulges.

\section{Discussion and further observations}  
Through a simple definition of infrared surface brightness based on the magnitude 
in a small fixed aperture, we have identified a sample of infrared LSB galaxies. 
This definition has been shown to work well at selecting LSB galaxies in the Coma 
cluster and SA57 region (Jarrett et al. 1998), and we find that our sample has low 
infrared surface brightness levels compared to optically-selected 
samples of HSBs and LSBs detected by 2MASS.

We had initially expected an infrared-selected sample of LSB galaxies to have 
redder colours than optical samples, but instead it appears that by requiring 
$\mu_{K5}>18$ \masq, we may have succeeded primarily in eliminating galaxies 
with significant bulge contributions. This already suggests that there is not 
a significant red population of LSB galaxies, although we note that the Large 
and Small 2MASS samples span a wide range of colours, particularly as measured 
in $B_{T_{c}}-K_T$.

To better understand the infrared LSB galaxies' properties we have carried out a 
multi-wavelength study of subsets of manageable size. These results will be 
described in subsequent papers (II-V).
For observations in the 21-cm \HI\ line we selected two subsets: one of 367
Large, Small and Faint 2MASS LSBs for observations at Arecibo and another of 334
Large 2MASS LSBs for observations at \nan, taking into account the difference 
in sensitivity and sky coverage of the two telescopes. The Arecibo and \nan\ data 
will be presented in, respectively, Papers II and III of the present series. 
We have obtained CCD surface photometry of 35 objects in the $B$, $V$, $R$ and 
$I$-bands using the 1.5-meter telescope at the San Pedro Mart{\'{i}}r 
Observatory in Mexico. These surface photometry data will be presented in Paper IV.
Finally, in Paper V we will further analyse the infrared LSBs to better understand, 
e.g. their properties as a function of their colours, investigating whether there 
is a significant population of red LSB galaxies, and to see if these objects 
follow the `standard' Tully-Fisher relation of HSB galaxies. In a related paper 
(Boissier et al. 2003), we present models of the evolution of samples of LSBs, 
including the 2MASS objects for which we obtained optical surface photometry.

\acknowledgements{ 
We want to thank the referee, Dr. N. Bergvall, for his comments.
This publication makes use of data products from the Two Micron All Sky Survey, 
which is a joint project of the University of Massachusetts and the Infrared 
Processing and Analysis Center, funded by the National Aeronautics and Space 
Administration and the National Science Foundation. This research also 
has made use of the Lyon-Meudon Extragalactic Database (LEDA), 
recently incorporated in HyperLeda, the NASA/IPAC 
Extragalactic Database (NED) which is operated by the Jet Propulsion Laboratory, 
California Institute of Technology, under contract with the National Aeronautics 
and Space Administration and the Aladin database, operated at the CDS, Strasbourg, 
France. We acknowledge financial support from CNRS/NSF collaboration grant No. 10637.
}

\newpage

 \begin{table*}
 \centering
 \bigskip
 {\scriptsize
\begin{tabular}{lrrllll}
 \multicolumn{7}{l}{{\bf Table 1a.} Large objects }\\
 \smallskip \\
 \hline
 \vspace{-2 mm} \\
 No & 2MASS &  PGC & Other & $\mu_{K5}$ & $\mu_{B_{25}}$ & Obs. \\
 \vspace{-2 mm} \\
 \hline
 \vspace{-2 mm} \\
L1P  &     0006240+465851   &      2290356   &                              &    18.02  &    24.77 & \\                              
L2P  &     0007208--280714   &         0571   &                              &    18.22  &    24.32 & \\
\end{tabular}
 }
 \normalsize
 \end{table*}
 \begin{table*}
 \centering
 \bigskip
 {\scriptsize
 \begin{tabular}{llrllll}
 \multicolumn{7}{l}{{\bf Table 1b.} Large objects - {\it continued} }\\
 \smallskip \\
 \hline
 \vspace{-2 mm} \\
 No & 2MASS &  PGC & Other & $\mu_{K5}$ & $\mu_{B_{25}}$ & Obs.. \\
 \vspace{-2 mm} \\
 \hline
 \vspace{-2 mm} \\
L81O &     0155396+370219   &         7189   &        CGCG 522-068          &   18.05  &   23.27  & a\\                              
L82O &     0156210+372707   &         7254   &        UGC 1405             &    18.85  &   24.06  & {\bf {O}}\\
\end{tabular}
 }
 \normalsize
 \end{table*}
 \begin{table*}
 \centering
 \bigskip
 {\scriptsize
 \begin{tabular}{llrllll}
 \multicolumn{7}{l}{{\bf Table 1c.} Large objects - {\it continued} }\\
 \smallskip \\
 \hline
 \vspace{-2 mm} \\
 No & 2MASS &  PGC & Other & $\mu_{K5}$ & $\mu_{B_{25}}$ & Obs.. \\
 \vspace{-2 mm} \\
 \hline
 \vspace{-2 mm} \\
L161O &     0306290--094332   &        11656   &        IC 1880               &    18.79  &   24.45  & {\bf {N}}\\   
L162O &     0306403--325147   &        11665   &        IC 1885               &    18.23  &   23.38  & \\
\end{tabular}
 }
 \normalsize
 \end{table*}
 \begin{table*}
 \centering
 \bigskip
 {\scriptsize
 \begin{tabular}{llrllll}
 \multicolumn{7}{l}{{\bf Table 1d.} Large objects - {\it continued} }\\
 \smallskip \\
 \hline
 \vspace{-2 mm} \\
 No & 2MASS &  PGC & Other & $\mu_{K5}$ & $\mu_{B_{25}}$  & Obs.. \\
 \vspace{-2 mm} \\
 \hline
 \vspace{-2 mm} \\
L241O &     0435219--085619   &       997298   &        FGC 494              &    18.04  &    25.88 & \\                              
L242O &     0436267--205043   &        15629   &        ESO 551-G18        &    18.06  &   22.77  & \\
\end{tabular}
 }
 \normalsize
 \end{table*}
 \begin{table*}
 \centering
 \bigskip
 {\scriptsize
 \begin{tabular}{llrllll}
 \multicolumn{7}{l}{{\bf Table 1e.} Large objects - {\it continued} }\\
 \smallskip \\
 \hline
 \vspace{-2 mm} \\
 No & 2MASS &  PGC & Other & $\mu_{K5}$ & $\mu_{B_{25}}$ & Obs. \\
 \vspace{-2 mm} \\
 \hline
 \vspace{-2 mm} \\
L321O &     0740336+341348   &        21555   &        IC 2203               &    18.19  &   23.645 & {\bf {O}}\\                              
L322O &     0740350+260805   &        21549   &        CGCG 117-064          &    18.28  &   23.813 & {\bf {O}}\\
\end{tabular}
 }
 \normalsize
 \end{table*}

 \begin{table*}
 \centering
 \bigskip
 {\scriptsize
 \begin{tabular}{llrllll}
 \multicolumn{7}{l}{{\bf Table 1f.} Large objects - {\it continued} }\\
 \smallskip \\
 \hline
 \vspace{-2 mm} \\
 No & 2MASS &  PGC & Other & $\mu_{K5}$ & $\mu_{B_{25}}$ & Obs. \\
 \vspace{-2 mm} \\
 \hline
 \vspace{-2 mm} \\
L401P &     1010413--074003   &      1015078   &                              &    18.82  &    23.48 & {\bf {N}}\\                              
L402O &     1014216+220728   &        29865   &        UGC 5524             &    18.48  &   23.94  & {\bf {O}}\\
\end{tabular}
 }
 \normalsize
 \end{table*}
 \begin{table*}
 \centering
 \bigskip
 {\scriptsize
 \begin{tabular}{llrllll}
 \multicolumn{7}{l}{{\bf Table 1g.} Large objects - {\it continued} }\\
 \smallskip \\
 \hline
 \vspace{-2 mm} \\
 No & 2MASS &  PGC & Other & $\mu_{K5}$ & $\mu_{B_{25}}$ & Obs. \\
 \vspace{-2 mm} \\
 \hline
 \vspace{-2 mm} \\
L481O &     1156249--121000   &        37518   &        MCG -02-31-002        &    18.54  &    23.46 & {\bf {N}}\\                              
L482O &     1157489+251615   &        37629   &        CGCG 127-114E         &    18.01  &   23.34  & {\bf {O}}\\
\end{tabular}
 }
 \normalsize
 \end{table*}
 \begin{table*}
 \centering
 \bigskip
 {\scriptsize
 \begin{tabular}{llrllll}
 \multicolumn{7}{l}{{\bf Table 1h.} Large objects - {\it continued} }\\
 \smallskip \\
 \hline
 \vspace{-2 mm} \\
 No & 2MASS &  PGC & Other & $\mu_{K5}$ & $\mu_{B_{25}}$ & Obs. \\
 \vspace{-2 mm} \\
 \hline
 \vspace{-2 mm} \\
L561P &     1309167--165311   &       892101   &                              &    18.89  &    24.69 & {\bf {N}}\\                              
L562O &     1309411--010253   &        45644   &        UGC 8238             &    18.27  &    23.49 & {\bf {O}}\\
\end{tabular}
 }
 \normalsize
 \end{table*}
 \begin{table*}
 \centering
 \bigskip
 {\scriptsize
 \begin{tabular}{llrllll}
 \multicolumn{7}{l}{{\bf Table 1i.} Large objects - {\it continued} }\\
 \smallskip \\
 \hline
 \vspace{-2 mm} \\
 No & 2MASS &  PGC & Other & $\mu_{K5}$ & $\mu_{B_{25}}$ & Obs. \\
 \vspace{-2 mm} \\
 \hline
 \vspace{-2 mm} \\
L641O &     1446510--315546   &        52775   &        ESO 447-G35        &    19.04  &   23.42  & {\bf {N}}\\                              
L642O &     1447354--180414   &        52827   &        ESO 580-G30        &    18.27  &   23.57  & {\bf {O}}\\
\end{tabular}
 }
 \normalsize
 \end{table*}
 \begin{table*}
 \centering
 \bigskip
 {\scriptsize
 \begin{tabular}{llrllll}
 \multicolumn{7}{l}{{\bf Table 1j.} Large objects - {\it continued} }\\
 \smallskip \\
 \hline
 \vspace{-2 mm} \\
 No & 2MASS &  PGC & Other & $\mu_{K5}$ & $\mu_{B_{25}}$ & Obs. \\
 \vspace{-2 mm} \\
 \hline
 \vspace{-2 mm} \\
L721O &     1831460+310523   &        61994   &        UGC 11264             &    18.05  &   23.43  &   \\                              
L722O &     1842201+460618   &        62282   &        CGCG 255-010          &    18.72  &   23.92  & {\bf {N}}\\
\end{tabular}
 }
 \normalsize
 \end{table*}
 \begin{table*}
 \centering
 \bigskip
 {\scriptsize
 \begin{tabular}{llrllll}
 \multicolumn{7}{l}{{\bf Table 1k.} Large objects - {\it continued} }\\
 \smallskip \\
 \hline
 \vspace{-2 mm} \\
 No & 2MASS &  PGC & Other & $\mu_{K5}$ & $\mu_{B_{25}}$ & Obs. \\
 \vspace{-2 mm} \\
 \hline
 \vspace{-2 mm} \\
L801O &     2258011--034609   &        70127   &        MCG -01-58-009        &    18.95  &   22.65  & \\    
L802P &     2258027+322757   &      2800849   &                              &    18.15  &          & {\bf {O}}, a\\
\end{tabular}
 }
 \normalsize
 \end{table*}

 \begin{table*}
 \centering
 \bigskip
 {\scriptsize
 \begin{tabular}{llrllll}
 \multicolumn{7}{l}{{\bf Table 2a.} Small objects }\\
 \smallskip \\
 \hline
 \vspace{-2 mm} \\
 No & 2MASS &  PGC & Other & $\mu_{K5}$ & $\mu_{B_{25}}$ & Obs. \\
 \vspace{-2 mm} \\
 \hline
 \vspace{-2 mm} \\
S1P &     0000033--104315   &       974615   &                                            &    18.26  &    24.99 & \\
S2N &     0000179+513620   &                &                                            &    18.73  &          & \\
\end{tabular}
 }
 \normalsize
 \end{table*}
 \begin{table*}
 \centering
 \bigskip
 {\scriptsize
 \begin{tabular}{llrllll}
 \multicolumn{7}{l}{{\bf Table 2b.} Small objects - {\it continued} }\\
 \smallskip \\
 \hline
 \vspace{-2 mm} \\
 No & 2MASS &  PGC & Other & $\mu_{K5}$ & $\mu_{B_{25}}$ & Obs. \\
 \vspace{-2 mm} \\
 \hline
 \vspace{-2 mm} \\
S81O &     0029264+172905   &         1803   &  MCG +03-02-014                            &    19.30  &    24.19 & a\\
S82O &     0029331--150549   &         1812   &  MCG -03-02-017                            &    19.16  &    23.26 & \\
\end{tabular}
 }
 \normalsize
 \end{table*}
 \begin{table*}
 \centering
 \bigskip
 {\scriptsize
 \begin{tabular}{llrllll}
 \multicolumn{7}{l}{{\bf Table 2c.} Small objects - {\it continued} }\\
 \smallskip \\
 \hline
 \vspace{-2 mm} \\
 No & 2MASS &  PGC & Other & $\mu_{K5}$ & $\mu_{B_{25}}$ & Obs. \\
 \vspace{-2 mm} \\
 \hline
 \vspace{-2 mm} \\
S161P &     0104249+211110   &      1644665   &                                            &    18.46  &    24.39 & a \\
S162P &     0105186+333615   &      2035469   &                                            &    18.37  &    23.98 & \\
\end{tabular}
 }
 \normalsize
 \end{table*}
 \begin{table*}
 \centering
 \bigskip
 {\scriptsize
 \begin{tabular}{llrllll}
 \multicolumn{7}{l}{{\bf Table 2d.} Small objects - {\it continued} }\\
 \smallskip \\
 \hline
 \vspace{-2 mm} \\
 No & 2MASS &  PGC & Other & $\mu_{K5}$ & $\mu_{B_{25}}$ & Obs. \\
 \vspace{-2 mm} \\
 \hline
 \vspace{-2 mm} \\
S241O &     0130201--121433   &         5606   &   MCG -02-04-061                           &    18.01  &    23.78 & \\
S242O &     0131238--123727   &         5659   &   IC 128                                  &    18.58  &    23.33 & \\
\end{tabular}
 }
 \normalsize
 \end{table*}
 \begin{table*}
 \centering
 \bigskip
 {\scriptsize
 \begin{tabular}{llrllll}
 \multicolumn{7}{l}{{\bf Table 2e.} Small objects - {\it continued} }\\
 \smallskip \\
 \hline
 \vspace{-2 mm} \\
 No & 2MASS &  PGC & Other & $\mu_{K5}$ & $\mu_{B_{25}}$ & Obs. \\
 \vspace{-2 mm} \\
 \hline
 \vspace{-2 mm} \\
S321P &     0202002+202546   &      1627043   &                                            &    18.47  &    24.33 & a \\
S322O &     0202320+445748   &         7767   &   UGC 1516                                &    18.38  &    23.63 & \\
\end{tabular}
 }
 \normalsize
 \end{table*}

 \begin{table*}
 \centering
 \bigskip
 {\scriptsize
 \begin{tabular}{llrllll}
 \multicolumn{7}{l}{{\bf Table 2f.} Small objects - {\it continued} }\\
 \smallskip \\
 \hline
 \vspace{-2 mm} \\
 No & 2MASS &  PGC & Other & $\mu_{K5}$ & $\mu_{B_{25}}$ & Obs.. \\
 \vspace{-2 mm} \\
 \hline
 \vspace{-2 mm} \\
S401P &     0232496+230820   &      1682146   &                                            &    18.44  &    23.95 & a \\
S402O &     0232538+231938   &         9707   &    UGC 2020                               &    18.53  &    23.56 & {\bf {O}}\\
\end{tabular}
 }
 \normalsize
 \end{table*}
 \begin{table*}
 \centering
 \bigskip
 {\scriptsize
 \begin{tabular}{llrllll}
 \multicolumn{7}{l}{{\bf Table 2g.} Small objects - {\it continued} }\\
 \smallskip \\
 \hline
 \vspace{-2 mm} \\
 No & 2MASS &  PGC & Other &$\mu_{K5}$ & $\mu_{B_{25}}$ & Obs.. \\
 \vspace{-2 mm} \\
 \hline
 \vspace{-2 mm} \\
S481P &     0305300--023502   &        11618   &                                            &    19.06  &    24.02 & {\bf {O}}\\
S482O &     0305365+413408   &       213161   &   UGC 2528                                &    18.57  &    25.25 & \\
\end{tabular}
 }
 \normalsize
 \end{table*}
 \begin{table*}
 \centering
 \bigskip
 {\scriptsize
 \begin{tabular}{llrllll}
 \multicolumn{7}{l}{{\bf Table 2h.} Small objects - {\it continued} }\\
 \smallskip \\
 \hline
 \vspace{-2 mm} \\
 No & 2MASS &  PGC & Other &$\mu_{K5}$ & $\mu_{B_{25}}$ & Obs.. \\
 \vspace{-2 mm} \\
 \hline
 \vspace{-2 mm} \\
S561P &     0326355--012207   &       176180   &                                            &    18.33  &    23.70 & \\
S562P &     0326367--284021   &       738890   &                                            &    18.01  &    24.41 & \\
\end{tabular}
 }
 \normalsize
 \end{table*}
 \begin{table*}
 \centering
 \bigskip
 {\scriptsize
 \begin{tabular}{llrllll}
 \multicolumn{7}{l}{{\bf Table 2i.} Small objects - {\it continued} }\\
 \smallskip \\
 \hline
 \vspace{-2 mm} \\
 No & 2MASS &  PGC & Other &$\mu_{K5}$ & $\mu_{B_{25}}$ & Obs.. \\
 \vspace{-2 mm} \\
 \hline
 \vspace{-2 mm} \\
S641P &     0357026--164640   &       893654   &                                            &    18.23  &    23.9  & \\
S642P &     0357115--272958   &       754047   &                                            &    18.15  &    24.46 & \\
\end{tabular}
 }
 \normalsize
 \end{table*}
 \begin{table*}
 \centering
 \bigskip
 {\scriptsize
 \begin{tabular}{llrllll}
 \multicolumn{7}{l}{{\bf Table 2j.} Small objects - {\it continued} }\\
 \smallskip \\
 \hline
 \vspace{-2 mm} \\
 No & 2MASS &  PGC & Other & $\mu_{K5}$ & $\mu_{B_{25}}$ & Obs. \\
 \vspace{-2 mm} \\
 \hline
 \vspace{-2 mm} \\
(S721N) &     (0424456+215937)   &                &                                            &    18.50  &          & a\\
S722P &     0425313--251205   &       781582   &                                            &    18.07  &    24.25 & \\
\end{tabular}
 }
 \normalsize
 \end{table*}

 \begin{table*}
 \centering
 \bigskip
 {\scriptsize
 \begin{tabular}{llrllll}
 \multicolumn{7}{l}{{\bf Table 2k.} Small objects - {\it continued} }\\
 \smallskip \\
 \hline
 \vspace{-2 mm} \\
 No & 2MASS &  PGC & Other & $\mu_{K5}$ & $\mu_{B_{25}}$ & Obs. \\
 \vspace{-2 mm} \\
 \hline
 \vspace{-2 mm} \\
S801P &     0455166--195955   &       847586   &                                            &    18.85  &    23.64 & \\
S802O &     0455167--224338   &        16334   &     ESO 485-G24                         &    18.45  &    23.27 & \\
\end{tabular}
 }
 \normalsize
 \end{table*}
 \begin{table*}
 \centering
 \bigskip
 {\scriptsize
 \begin{tabular}{llrllll}
 \multicolumn{7}{l}{{\bf Table 2l.} Small objects - {\it continued} }\\
 \smallskip \\
 \hline
 \vspace{-2 mm} \\
 No & 2MASS &  PGC & Other & $\mu_{K5}$ & $\mu_{B_{25}}$ & Obs.. \\
 \vspace{-2 mm} \\
 \hline
 \vspace{-2 mm} \\
S881P &     0526376--265141   &       762993   &                                            &    19.00  &    23.14 & \\
S882P &     0527141--150430   &       178449   &                                            &    18.12  &    23.33 & \\
\end{tabular}
 }
 \normalsize
 \end{table*}
 \begin{table*}
 \centering
 \bigskip
 {\scriptsize
 \begin{tabular}{llrllll}
 \multicolumn{7}{l}{{\bf Table 2m.} Small objects - {\it continued} }\\
 \smallskip \\
 \hline
 \vspace{-2 mm} \\
 No & 2MASS &  PGC & Other & $\mu_{K5}$ & $\mu_{B_{25}}$ & Obs.. \\
 \vspace{-2 mm} \\
 \hline
 \vspace{-2 mm} \\
S961P &     0634292--342401   &       662842   &                                            &    18.35  &    23.91 & \\
S962P &     0634533--221855   &       816078   &                                            &    18.63  &    24.23 & \\
\end{tabular}
 }
 \normalsize
 \end{table*}
 \begin{table*}
 \centering
 \bigskip
 {\scriptsize
 \begin{tabular}{llrllll}
 \multicolumn{7}{l}{{\bf Table 2n.} Small objects - {\it continued} }\\
 \smallskip \\
 \hline
 \vspace{-2 mm} \\
 No & 2MASS &  PGC & Other & $\mu_{K5}$ & $\mu_{B_{25}}$ & Obs.. \\
 \vspace{-2 mm} \\
 \hline
 \vspace{-2 mm} \\
S1041P &     0731059+474555   &      2304003   &                                            &    18.79  &    24.85 & \\
S1042P &     0731070+412539   &      2180950   &                                            &    18.10  &    24.39 & \\
\end{tabular}
 }
 \normalsize
 \end{table*}
 \begin{table*}
 \centering
 \bigskip
 {\scriptsize
 \begin{tabular}{llrllll}
 \multicolumn{7}{l}{{\bf Table 2o.} Small objects - {\it continued} }\\
 \smallskip \\
 \hline
 \vspace{-2 mm} \\
 No & 2MASS &  PGC & Other & $\mu_{K5}$ & $\mu_{B_{25}}$ & Obs. \\
 \vspace{-2 mm} \\
 \hline
 \vspace{-2 mm} \\
S1121P &     0815524--105034   &       973066   &                                            &    18.31  &    25.15 & \\
S1122N &     0816236+222258   &                &                                            &    18.02  &          & a\\
\end{tabular}
 }
 \normalsize
 \end{table*}

 \begin{table*}
 \centering
 \bigskip
 {\scriptsize
 \begin{tabular}{llrllll}
 \multicolumn{7}{l}{{\bf Table 2p.} Small objects - {\it continued} }\\
 \smallskip \\
 \hline
 \vspace{-2 mm} \\
 No & 2MASS &  PGC & Other & $\mu_{K_{5}}$ & $\mu_{B_{25}}$ & Obs. \\
 \vspace{-2 mm} \\
 \hline
 \vspace{-2 mm} \\
S1201P &     0851505+192933   &      1594342   &                                            &    18.71  &    23.81 & \\
S1202P &     0852380--052522   &      1044922   &                                            &    18.24  &    25.18 & \\
\end{tabular}
 }
 \normalsize
 \end{table*}
 \begin{table*}
 \centering
 \bigskip
 {\scriptsize
 \begin{tabular}{llrllll}
 \multicolumn{7}{l}{{\bf Table 2q.} Small objects - {\it continued} }\\
 \smallskip \\
 \hline
 \vspace{-2 mm} \\
 No & 2MASS &  PGC & Other & $\mu_{K_{5}}$ & $\mu_{B_{25}}$ & Obs. \\
 \vspace{-2 mm} \\
 \hline
 \vspace{-2 mm} \\
S1281P &     0922183+403404   &      2167756   &                                            &    18.27  &    23.98 & \\
S1282P &     0922268+230544   &      1681359   &                                            &    18.17  &    23.44 & \\
\end{tabular}
 }
 \normalsize
 \end{table*}
 \begin{table*}
 \centering
 \bigskip
 {\scriptsize
 \begin{tabular}{llrllll}
 \multicolumn{7}{l}{{\bf Table 2r.} Small objects - {\it continued} }\\
 \smallskip \\
 \hline
 \vspace{-2 mm} \\
 No & 2MASS &  PGC & Other & $\mu_{K_{5}}$ & $\mu_{B_{25}}$ & Obs. \\
 \vspace{-2 mm} \\
 \hline
 \vspace{-2 mm} \\
S1361P &     0957323+362028   &      2079003   &                                            &    18.35  &    24.84 & \\
S1362P &     0957582--102515   &       978434   &                                            &    18.49  &    24.26 & \\
\end{tabular}
 }
 \normalsize
 \end{table*}
 \begin{table*}
 \centering
 \bigskip
 {\scriptsize
 \begin{tabular}{llrllll}
 \multicolumn{7}{l}{{\bf Table 2s.} Small objects - {\it continued} }\\
 \smallskip \\
 \hline
 \vspace{-2 mm} \\
 No & 2MASS &  PGC & Other & $\mu_{K_{5}}$ & $\mu_{B_{25}}$ & Obs. \\
 \vspace{-2 mm} \\
 \hline
 \vspace{-2 mm} \\
S1441P &     1033103+215105   &      1658590   &                                            &    19.21  &    24.07 & \\
S1442P &     1034434+202138   &      1624987   &                                            &    18.07  &    25.49 & a\\
\end{tabular}
 }
 \normalsize
 \end{table*}
 \begin{table*}
 \centering
 \bigskip
 {\scriptsize
 \begin{tabular}{llrllll}
 \multicolumn{7}{l}{{\bf Table 2t.} Small objects - {\it continued} }\\
 \smallskip \\
 \hline
 \vspace{-2 mm} \\
 No & 2MASS &  PGC & Other & $\mu_{K_{5}}$ & $\mu_{B_{25}}$ & Obs. \\
 \vspace{-2 mm} \\
 \hline
 \vspace{-2 mm} \\
S1521P &     1111564--133937   &       182495   &                                            &    18.26  &    23.01 & \\
S1522P &     1112086+384557   &      2135997   &                                            &    18.23  &    23.67 & \\
\end{tabular}
 }
 \normalsize
 \end{table*}

 \begin{table*}
 \centering
 \bigskip
 {\scriptsize
 \begin{tabular}{llrllll}
 \multicolumn{7}{l}{{\bf Table 2u.} Small objects - {\it continued} }\\
 \smallskip \\
 \hline
 \vspace{-2 mm} \\
 No & 2MASS &  PGC & Other & $\mu_{K5}$ & $\mu_{B_{25}}$ & Obs. \\
 \vspace{-2 mm} \\
 \hline
 \vspace{-2 mm} \\
S1601P &     1142550+115847   &      1401817   &                                            &    18.11  &    24.32 & \\
S1602N &     1142593--251801   &                &                                            &    18.20  &          & \\
\end{tabular}
 }
 \normalsize
 \end{table*}
 \begin{table*}
 \centering
 \bigskip
 {\scriptsize
 \begin{tabular}{llrllll}
 \multicolumn{7}{l}{{\bf Table 2v.} Small objects - {\it continued} }\\
 \smallskip \\
 \hline
 \vspace{-2 mm} \\
 No & 2MASS &  PGC & Other & $\mu_{K5}$ & $\mu_{B_{25}}$ & Obs. \\
 \vspace{-2 mm} \\
 \hline
 \vspace{-2 mm} \\
S1681O &     1212032+145416   &        38868   &    VCC 040                                &    19.23  &    23.62 & {\bf {O}}\\
S1682N &     1212065--261943   &                &                                            &    18.58  &          & \\
\end{tabular}
 }
 \normalsize
 \end{table*}
 \begin{table*}
 \centering
 \bigskip
 {\scriptsize
 \begin{tabular}{llrllll}
 \multicolumn{7}{l}{{\bf Table 2w.} Small objects - {\it continued} }\\
 \smallskip \\
 \hline
 \vspace{-2 mm} \\
 No & 2MASS &  PGC & Other & $\mu_{K5}$ & $\mu_{B_{25}}$ & Obs. \\
 \vspace{-2 mm} \\
 \hline
 \vspace{-2 mm} \\
S1761P &     1245551+181019   &      1552147   &                                            &    18.52  &    25.25 & \\
S1762O &     1246019--103813   &                &   NPM1G -10.0446                           &    18.30  &          & \\
\end{tabular}
 }
 \normalsize
 \end{table*}
 \begin{table*}
 \centering
 \bigskip
 {\scriptsize
 \begin{tabular}{llrllll}
 \multicolumn{7}{l}{{\bf Table 2x.} Small objects - {\it continued} }\\
 \smallskip \\
 \hline
 \vspace{-2 mm} \\
 No & 2MASS &  PGC & Other & $\mu_{K5}$ & $\mu_{B_{25}}$ & Obs. \\
 \vspace{-2 mm} \\
 \hline
 \vspace{-2 mm} \\
S1841P &     1310557+115228   &      1400130   &                                            &    18.70  &    24.72 & \\
S1842O &     1311371+320333   &      1978264   &    NGP9 F323-0051554                       &    18.06  &    23.64 & \\
\end{tabular}
 }
 \normalsize
 \end{table*}
 \begin{table*}
 \centering
 \bigskip
 {\scriptsize
 \begin{tabular}{llrllll}
 \multicolumn{7}{l}{{\bf Table 2y.} Small objects - {\it continued} }\\
 \smallskip \\
 \hline
 \vspace{-2 mm} \\
 No & 2MASS &  PGC & Other & $\mu_{K5}$ & $\mu_{B_{25}}$ & Obs. \\
 \vspace{-2 mm} \\
 \hline
 \vspace{-2 mm} \\
S1921P &     1330566--010909   &      1125472   &                                            &    18.79  &    24.24 & \\
S1922O &     1331125+414746   &      2186982   &    IRAS F13290+4202                        &    18.88  &    24.79 & \\
\end{tabular}
 }
 \normalsize
 \end{table*}

 \begin{table*}
 \centering
 \bigskip
 {\scriptsize
 \begin{tabular}{llrllll}
 \multicolumn{7}{l}{{\bf Table 2z.} Small objects - {\it continued} }\\
 \smallskip \\
 \hline
 \vspace{-2 mm} \\
 No & 2MASS &  PGC & Other & $\mu_{K5}$ & $\mu_{B_{25}}$ & Obs. \\
 \vspace{-2 mm} \\
 \hline
 \vspace{-2 mm} \\
S2001P &     1401354--330322   &        89817   &                                            &    18.25  &    24.24 & \\
S2002P &     1401569--295634   &       723329   &                                            &    18.51  &    24.79 & \\
\end{tabular}
 }
 \normalsize
 \end{table*}
 \begin{table*}
 \centering
 \bigskip
 {\scriptsize
 \begin{tabular}{llrllll}
 \multicolumn{7}{l}{{\bf Table 2aa.} Small objects - {\it continued} }\\
 \smallskip \\
 \hline
 \vspace{-2 mm} \\
 No & 2MASS &  PGC & Other & $\mu_{K5}$ & $\mu_{B_{25}}$ & Obs. \\
 \vspace{-2 mm} \\
 \hline
 \vspace{-2 mm} \\
S2081N &     1444527--153240   &                &                                            &    18.59  &          & \\
S2082O &     1445068--031343   &      1074341   &    LCRS B144230.7-030107                   &    19.03  &    24.62 & \\
\end{tabular}
 }
 \normalsize
 \end{table*}
 \begin{table*}
 \centering
 \bigskip
 {\scriptsize
 \begin{tabular}{llrllll}
 \multicolumn{7}{l}{{\bf Table 2ab.} Small objects - {\it continued} }\\
 \smallskip \\
 \hline
 \vspace{-2 mm} \\
 No & 2MASS &  PGC & Other & $\mu_{K5}$ & $\mu_{B_{25}}$ & Obs. \\
 \vspace{-2 mm} \\
 \hline
 \vspace{-2 mm} \\
S2161P &     1534514+390924   &      2144450   &                                            &    18.31  &    23.85 & \\
S2162O &     1536073+593359   &        55561   &    NGC 5976A                               &    18.29  &    23.72 & \\
\end{tabular}
 }
 \normalsize
 \end{table*}
 \begin{table*}
 \centering
 \bigskip
 {\scriptsize
 \begin{tabular}{llrllll}
 \multicolumn{7}{l}{{\bf Table 2ac.} Small objects - {\it continued} }\\
 \smallskip \\
 \hline
 \vspace{-2 mm} \\
 No & 2MASS &  PGC & Other & $\mu_{K5}$ & $\mu_{B_{25}}$ & Obs. \\
 \vspace{-2 mm} \\
 \hline
 \vspace{-2 mm} \\
S2241O &     1626529+411513   &        58149   &     KUG 1625+413                           &    18.17  &    22.87 & \\
S2242P &     1627003+275548   &      1821778   &                                            &    18.26  &    24.15 & \\
\end{tabular}
 }
 \normalsize
 \end{table*}
 \begin{table*}
 \centering
 \bigskip
 {\scriptsize
 \begin{tabular}{llrllll}
 \multicolumn{7}{l}{{\bf Table 2ad.} Small objects - {\it continued} }\\
 \smallskip \\
 \hline
 \vspace{-2 mm} \\
 No & 2MASS &  PGC & Other & $\mu_{K5}$ & $\mu_{B_{25}}$ & Obs. \\
 \vspace{-2 mm} \\
 \hline
 \vspace{-2 mm} \\
S2321O &     1720327+174612   &        60104   &      CGCG  111-23                          &    19.06  &    24.28 & \\
S2322P &     1720599+133729   &      1435991   &                                            &    18.22  &    24.85 & \\
\end{tabular}
 }
 \normalsize
 \end{table*}

 \begin{table*}
 \centering
 \bigskip
 {\scriptsize
 \begin{tabular}{llrllll}
 \multicolumn{7}{l}{{\bf Table 2ae.} Small objects - {\it continued} }\\
 \smallskip \\
 \hline
 \vspace{-2 mm} \\
 No & 2MASS &  PGC & Other & $\mu_{K5}$ & $\mu_{B_{25}}$ & Obs. \\
 \vspace{-2 mm} \\
 \hline
 \vspace{-2 mm} \\
S2401P &     1840110+212933   &        62225   &       UGC 11314                            &    18.20  &    23.74 & \\ 
S2402P &     1840350+522029   &      2412573   &                                            &    18.70  &    23.67 & \\
\end{tabular}
 }
 \normalsize
 \end{table*}
 \begin{table*}
 \centering
 \bigskip
 {\scriptsize
 \begin{tabular}{llrllll}
 \multicolumn{7}{l}{{\bf Table 2af.} Small objects - {\it continued} }\\
 \smallskip \\
 \hline
 \vspace{-2 mm} \\
 No & 2MASS &  PGC & Other & $\mu_{K5}$ & $\mu_{B_{25}}$ & Obs. \\
 \vspace{-2 mm} \\
 \hline
 \vspace{-2 mm} \\
S2481P &     2012096--333319   &       674781   &                                            &    18.34  &    23.09 & \\
S2482P &     2013151--333732   &       673751   &                                            &    18.23  &    23.63 & \\
\end{tabular}
 }
 \normalsize
 \end{table*}
 \begin{table*}
 \centering
 \bigskip
 {\scriptsize
 \begin{tabular}{llrllll}
 \multicolumn{7}{l}{{\bf Table 2ag.} Small objects - {\it continued} }\\
 \smallskip \\
 \hline
 \vspace{-2 mm} \\
 No & 2MASS &  PGC & Other & $\mu_{K5}$ & $\mu_{B_{25}}$ & Obs. \\
 \vspace{-2 mm} \\
 \hline
 \vspace{-2 mm} \\
S2561P &     2120529--303539   &       715354   &                                            &    18.79  &    23.79 & \\
S2562P &     2120576+130851   &      1424009   &                                            &    18.49  &    23.74 & a\\
\end{tabular}
 }
 \normalsize
 \end{table*}
 \begin{table*}
 \centering
 \bigskip
 {\scriptsize
 \begin{tabular}{llrllll}
 \multicolumn{7}{l}{{\bf Table 2ah.} Small objects - {\it continued} }\\
 \smallskip \\
 \hline
 \vspace{-2 mm} \\
 No & 2MASS &  PGC & Other & $\mu_{K5}$ & $\mu_{B_{25}}$ & Obs. \\
 \vspace{-2 mm} \\
 \hline
 \vspace{-2 mm} \\
S2641N &     2216015+405206   &                &                                            &    18.18  &          & \\
S2642P &     2216140+162605   &      1508710   &                                            &    18.20  &    23.17 & \\
\end{tabular}
 }
 \normalsize
 \end{table*}
 \begin{table*}
 \centering
 \bigskip
 {\scriptsize
 \begin{tabular}{llrllll}
 \multicolumn{7}{l}{{\bf Table 2ai.} Small objects - {\it continued} }\\
 \smallskip \\
 \hline
 \vspace{-2 mm} \\
 No & 2MASS &  PGC & Other & $\mu_{K5}$ & $\mu_{B_{25}}$ & Obs. \\
 \vspace{-2 mm} \\
 \hline
 \vspace{-2 mm} \\
S2721P &     2251282--094544   &       986750   &                                            &    18.14  &    24.36 & \\
S2722P &     2251520--070620   &      1024334   &                                            &    18.20  &    23.68 & \\
\end{tabular}
 }
 \normalsize
 \end{table*}

 \begin{table*}
 \centering
 \bigskip
 {\scriptsize
 \begin{tabular}{llrllll}
 \multicolumn{7}{l}{{\bf Table 2aj.} Small objects - {\it continued} }\\
 \smallskip \\
 \hline
 \vspace{-2 mm} \\
 No & 2MASS &  PGC & Other & $\mu_{K5}$ & $\mu_{B_{25}}$ & Obs. \\
 \vspace{-2 mm} \\
 \hline
 \vspace{-2 mm} \\
S2801P &     2319183--020517   &      1102064   &                                            &    18.22  &    23.97 & \\
S2802P &     2319381+201321   &      1620517   &                                            &    18.50  &    24.17 & \\
\end{tabular}
 }
 \normalsize
 \end{table*}
 \begin{table*}
 \centering
 \bigskip
 {\scriptsize
 \begin{tabular}{llrllll}
 \multicolumn{7}{l}{{\bf Table 2ak.} Small objects - {\it continued} }\\
 \smallskip \\
 \hline
 \vspace{-2 mm}\\
 No & 2MASS &  PGC & Other & $\mu_{K5}$ & $\mu_{B_{25}}$ & Obs. \\
 \vspace{-2 mm} \\
 \hline
 \vspace{-2 mm} \\
S2881O &     2353111--283525   &        72741   &    ESO 471-G3                           &    19.54  &    23.96 & \\
S2882P &     2354411+121619   &      1406608   &                                            &    18.79  &    24.12 & \\
\end{tabular}
 }
 \normalsize
 \end{table*}

 \begin{table*}
 \centering
 \bigskip
 {\scriptsize
 \begin{tabular}{llrllll}
 \multicolumn{7}{l}{{\bf Table 3.} LCSB objects -- the Faint sources sample}\\
 \smallskip \\
 \hline
 \vspace{-2 mm} \\
 No & 2MASS &  PGC & Other & $\mu_{K5}$ & $\mu_{B_{25}}$ & Obs.\\
 \vspace{-2 mm} \\
 \hline
 \vspace{-2 mm} \\
(F1O) &   (0037501+172258) &  2256 & CGCG 457-023   & 19.4 & 23.6& {\bf {O}}, {\bf {A}}\\
(F2O) &   (0041167--051707) &  2470 & MCG -01-02-045 & 19.4 & 23.5 & {\bf {N}},P\\
\end{tabular}
 }
 \normalsize
 \end{table*}

\newpage


\begin{figure*}[!h] 
\centering
\includegraphics[width=15cm]{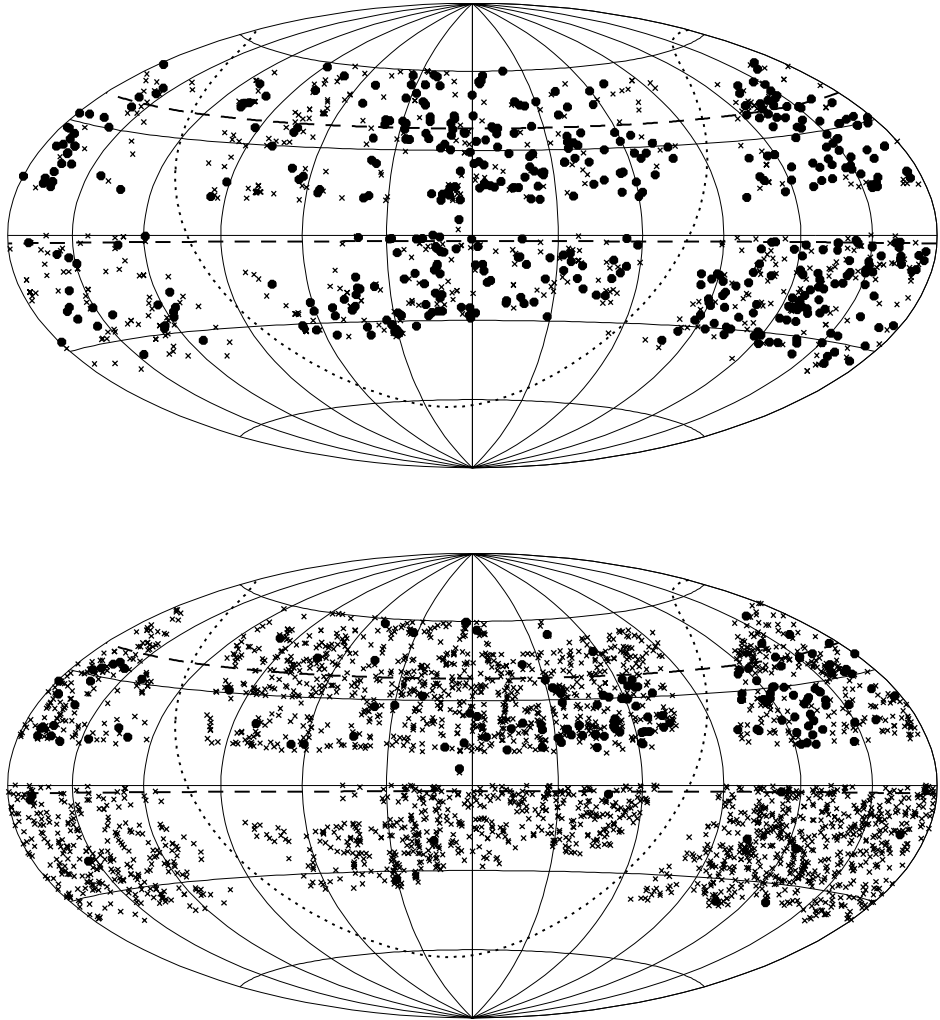}
\caption [] {The sky coverage of the selected sample of 2MASS LSB galaxies 
is illustrated by the distribution of the objects which were observed in \HI. 
The plot is in equatorial coordinates; the dotted line indicates the Galactic 
plane and the dashed lines the -2$^{\circ}$ to 38$^{\circ}$ declination range 
covered by the Arecibo radio telescope. All sources have declinations between
-39$^{\circ}$ and +60$^{\circ}$. 
The upper panel is for the Large sources 
($r_{K_{20}}>20''$), the lower for the Small sources ($20''>r_{K_{20}}>10''$). 
Dots indicate galaxies detected in \HI, crosses undetected objects.}
\end{figure*}

%
\begin{figure*} 
\centering
\includegraphics[width=15cm]{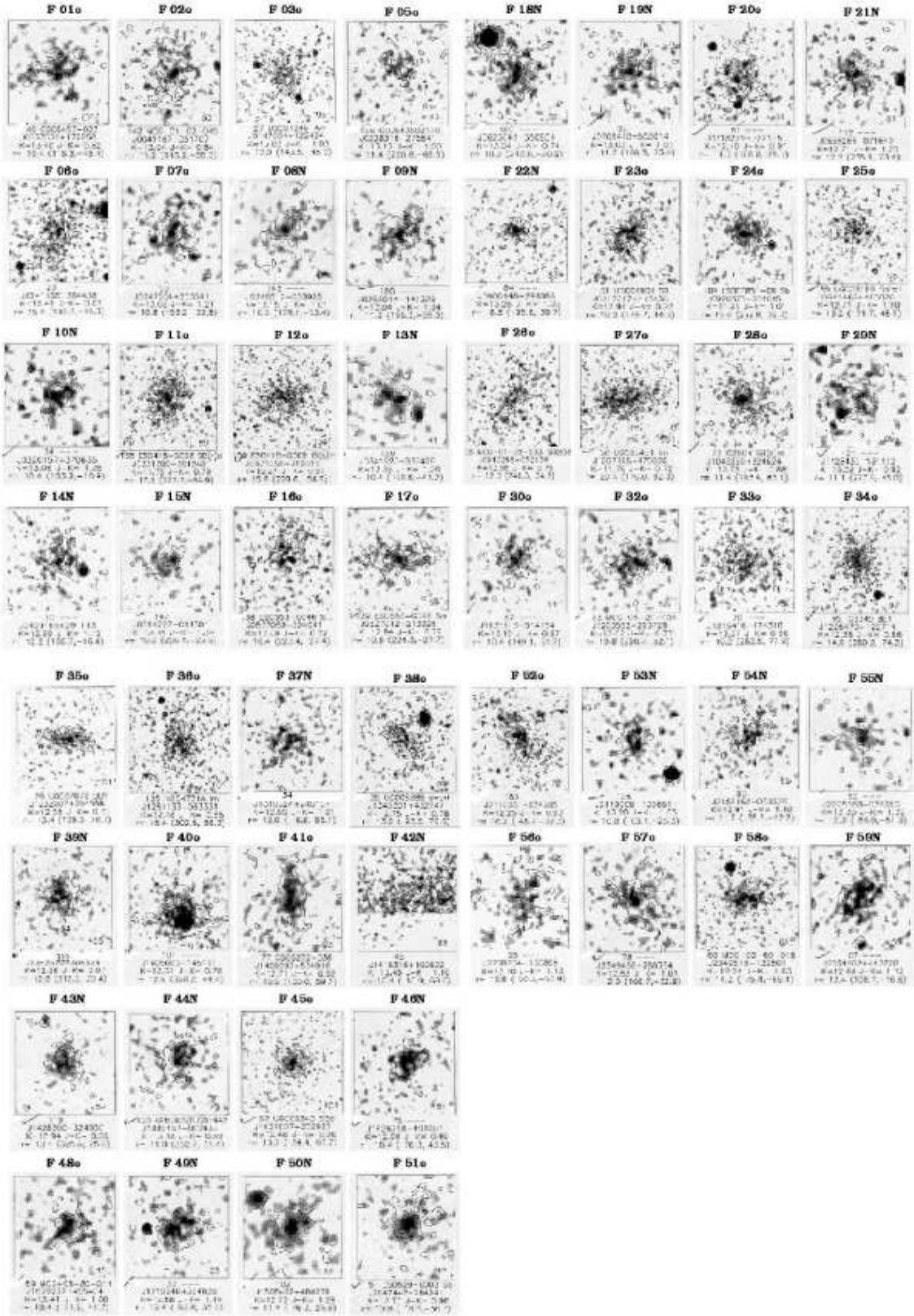}
\caption[]{{\bf a.-d.}\, 2MASS images of the selected LCSB sources: 
$J$-band greyscale images with superimposed $K_s$-band contours.
The size of each image (in arcsec) is indicated in its bottom right corner.
}
\end{figure*}

\begin{figure*} 
\centering
\includegraphics{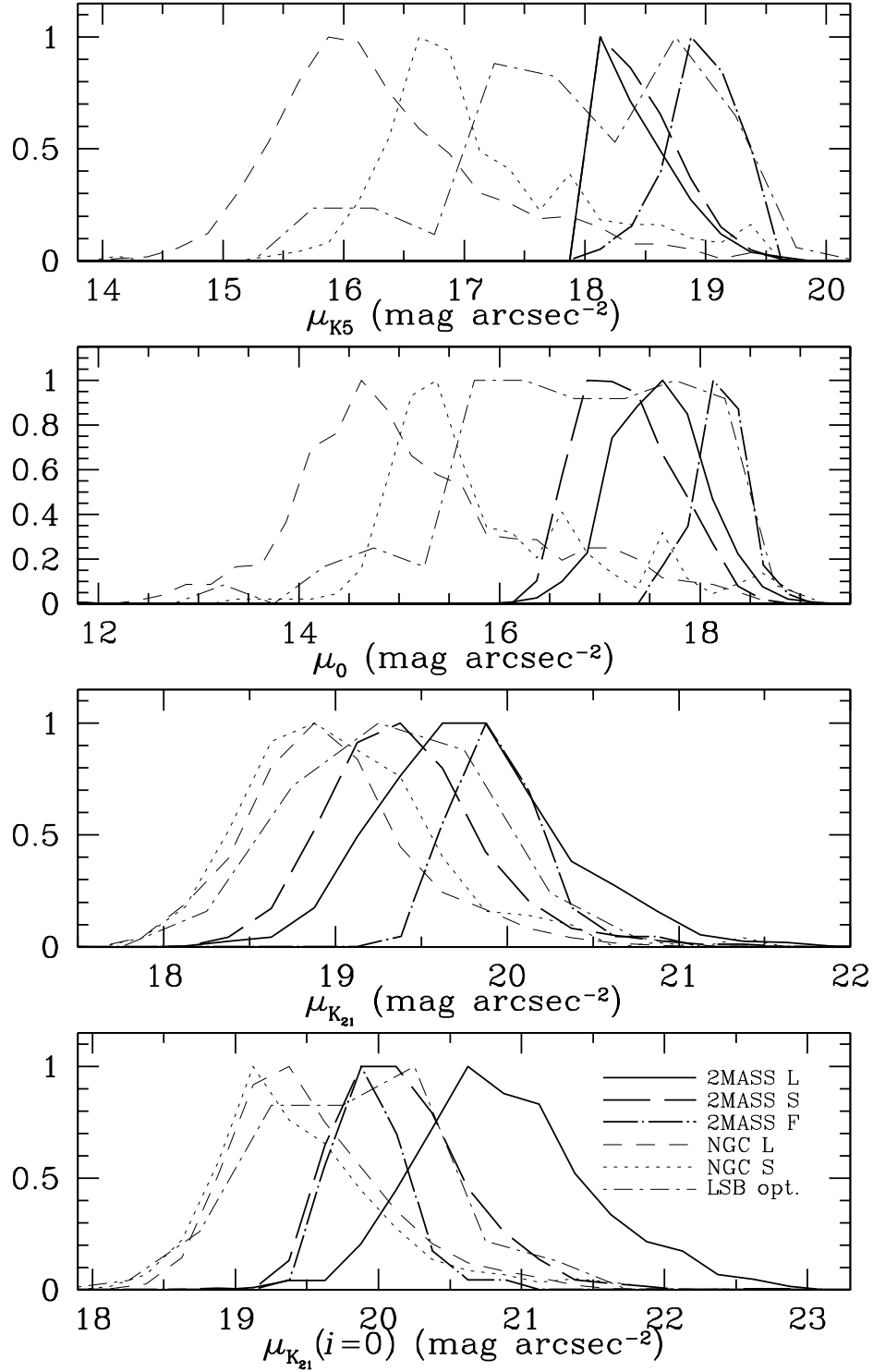}
\caption [] {Comparison between the normalized distributions of several measures of
infrared surface brightness in the $K_s$-band for the three sub-samples
of 2MASS LSB galaxies (Large/Small/Faint) with three other samples of galaxies,
one of optically selected LSBs and two of HSB galaxies from the NGC, as described 
in Section 4.1. 
The plotted parameters, which are described and discussed further in Section 4.2,
are: $\mu_{K5}$ is the  mean surface brightness measured within a circle of 5$''$ radius 
at the centre of each galaxy, $\mu_0$ is the highest surface brightness measured at the centre 
of the galaxy, and $\mu_{K_{21}}$ and $\mu_{K_{21}} (i=0)$ are, 
respectively, the average surface brightness measured within the 21 \masq\ isophotal 
contour in the  $K_s$-band and this value corrected to face-on $(i=0^{\circ})$.
}
\end{figure*}

\begin{figure*}
\includegraphics{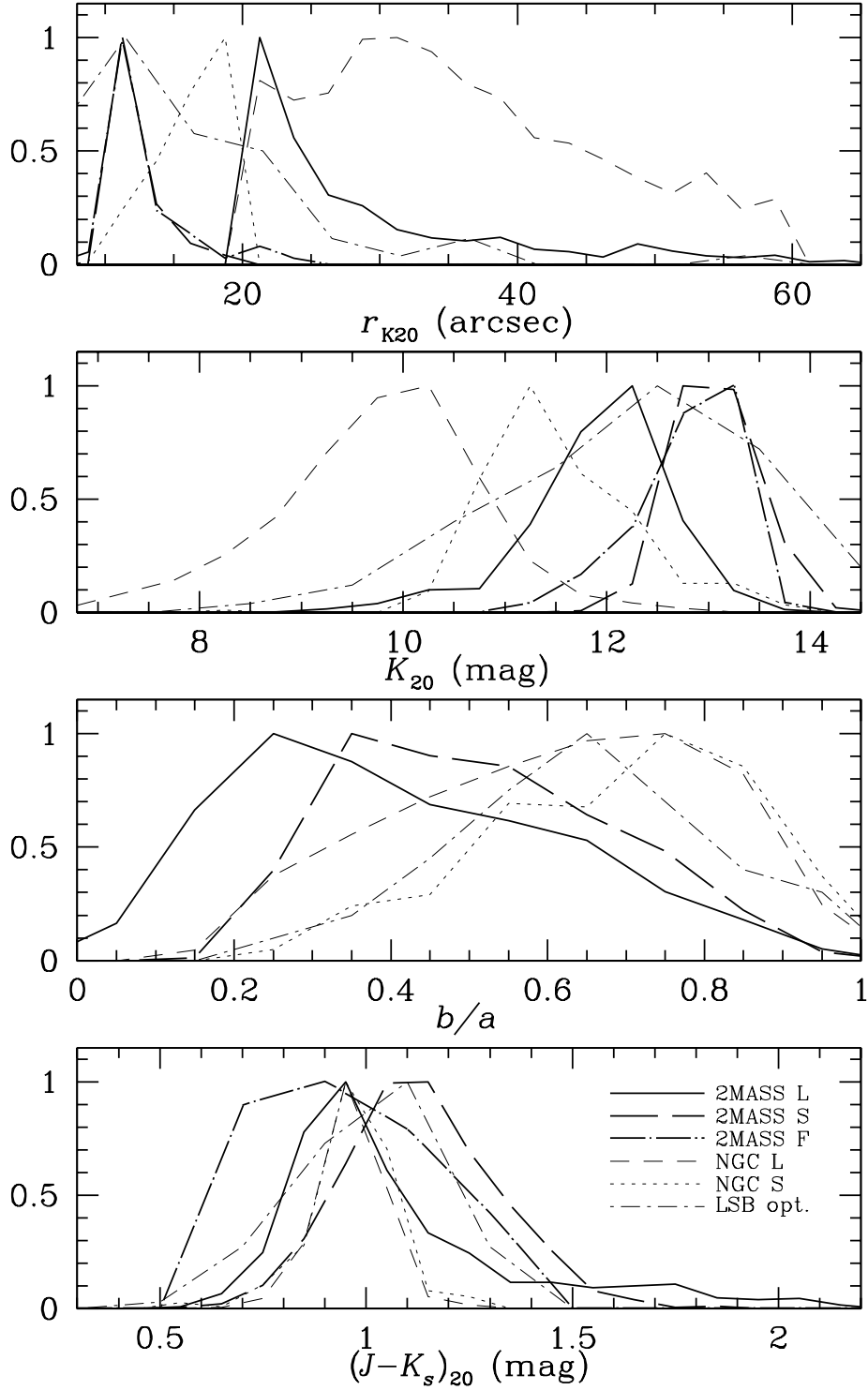}
\caption [] {As in Fig.~3, for a number of other infrared photometric parameters:
$r_{K_{20}}$ is the major-axis radius at the 20 \masq\ isophotal level in the $K_s$ band,
$K_{20}$ is the total magnitude within the 20 \masq\ elliptical isophote in the $K_s$ band,
$b/a$ is the minor-to-major axis ratio determined from an ellipse fit 
to the co-addition of the $J$-, $H$-, and $K_s$-band images, and
$(J-K)_{20}$ is the difference between the $J$- and $K_s$-band magnitudes measured 
within the $K_s=20$ \masq\ elliptical isophote; note that the left half of the NGC S and NGC L 
curves overlap in the bottom panel.
See Section 4.3 for further details on these parameters.
}
\end{figure*}

\begin{figure*}
\includegraphics{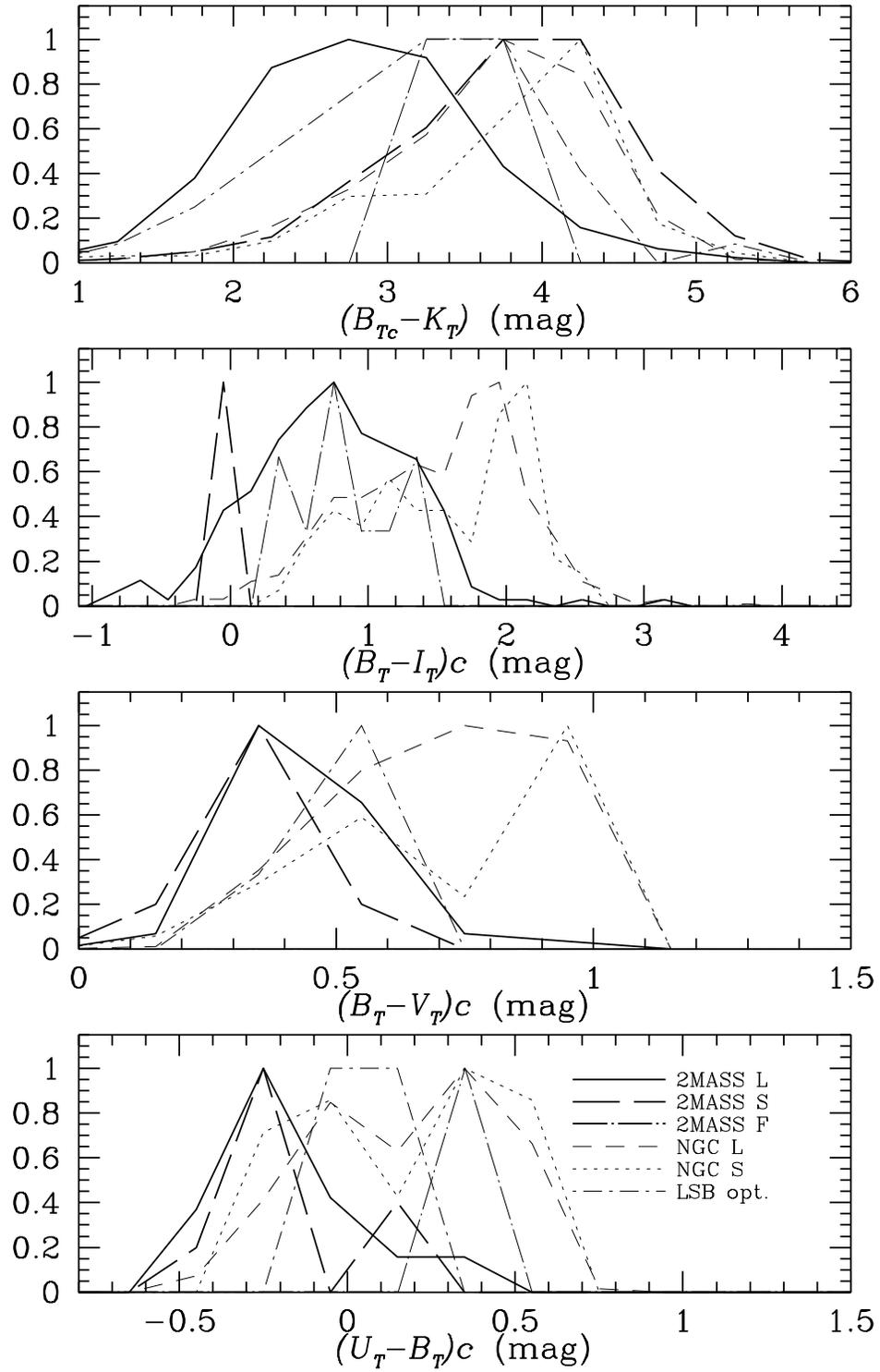}
\caption [] {As in Fig.~3 for a number of optical and
optical-infrared colours, based on total corrected extrapolated magnitudes from 2MASS measured in the
 $K_s$ band and from LEDA measured in the $U$, $B$, $V$ and $I$ bands. 
See Section 4.4 for a discussion of the plotted parameters.
}
\end{figure*}

\end{document}